\documentclass[aps,prx,twocolumn,longbibliography,amsmath,amsfonts,superscriptaddress,]{revtex4-2}
\usepackage[dvipdfmx]{graphicx}
\usepackage{dcolumn}
\usepackage{bm}
\usepackage{boxedminipage}
\usepackage{amsmath}
\usepackage{graphicx}
\usepackage{amsfonts}
\usepackage{feynmf}
\usepackage{mathrsfs}
\usepackage{listings}
\usepackage{algorithm}
\usepackage{algorithmic}
\usepackage{color}

\usepackage{hyperref}

\begin{document}


\title{Topological piezoelectric effect and parity anomaly in nodal line semimetals}


\author{Taiki Matsushita}
\affiliation{Department of Materials Engineering Science, Osaka University, Toyonaka, Osaka 560-8531, Japan}
\author{Satoshi Fujimoto}
\affiliation{Department of Materials Engineering Science, Osaka University, Toyonaka, Osaka 560-8531, Japan}
\author{Andreas P. Schnyder}
\affiliation{Max-Planck-Institute for Solid State Research, Heisenbergstrasse 1, D-70569 Stuttgart, Germany}

\date{\today}

\begin{abstract}
Lattice deformations act on the low-energy excitations of Dirac materials as effective axial vector fields.
This allows to directly detect  quantum anomalies of Dirac materials via the response to axial gauge fields. 
We investigate the parity anomaly in Dirac nodal line semimetals induced by lattice vibrations, and establish a topological piezoelectric effect; 
i.e., periodic lattice deformations generate topological Hall currents that
are transverse to the deformation field.
The currents induced by this piezoelectric effect are dissipationless
and their magnitude is completely determined by the length of the nodal ring, leading to a semi-quantized transport coefficient.
Our theoretical proposal can be exprimentally realized in various nodal line semimetals, such as CaAgP and Ca$_{_3}$P${_2}$.
\end{abstract}

\maketitle

\section{Introduction}
Over the last few years a number of new types of topological semimetals have been discovered~\cite{HosurWSM,armitagereview,WehlingDirac,Murakamiphase,Wantopological,burkovTSM,Nakadaedge}.
Among them are Weyl semimetals and Dirac semimetals with point nodes, around which the bands
have linear dispersion in all directions. 
The low-energy physics of these point node semimetals is described by relativistic field theories 
with quantum anomalies, i.e., by quantum field theories that break symmetries of the classical action.
For instance, two-dimensional Dirac materials, such as graphene, are described
by quantum field theories with parity anomalies,  
that break space-time inversion symmetry~\cite{Redlichparityanomaly,haldane_PRL_88}.
The low-energy theories of Weyl semimetals, on the other hand, 
exhibit chiral anomalies, which violate conservation of axial charge
~\cite{Nielsenanomaly,ZyuzinTQFT,GoswamiTQFT,Sonanomaly,BurkovTDSManomaly}.
The chiral anomaly in Weyl semimetals give rise to numerous experimental phenomena~\cite{yangAHE,FukushimaCME,VazifehCME,GoswamiNMR,HuangNMR},
for example, negative magnetoresistance, which has been observed in recent experiments~\cite{HuangNMR,Xiong413}.
Lattice strain, which generates axial magnetic fields, can also be used to probe the chiral anomaly~\cite{ilan_grushin_review_nat_2020,guineastrain,levystrain,Niggestrain,Liuaxialgauge,Grushinaxialgauge,Hughestorsion,Parrikartorsion,Sumiyoshiaxialgauge,Pikulinaxialgauge,kobayashitorsion,ishiharatorsion,Matsushitaaxilgauge,kurebayashimagneticdomain}.

At the same time, recent research has focused on Dirac materials with line nodes~\cite{ali_nodal_line_review_2018,yamakageCaAgAs,NeupaneARPES,takane2018CaAgAsARPS,XuARPESCaAgP,bianARPES,fengARPES,ominatoNLSM,ChanCa3P2}. 
These nodal-line semi\-metals (NLSMs) can be viewed as three-dimensional generalizations of graphene.
They exhibit Dirac band crossings along a one-dimensional  line in a three-dimensional Brillouin zone, with low-energy excitations
that are linearly dispersing in the two directions perpendicular to the band-crossing line. 
NLSMs possess a number of interesting properties, e.g., topological surface charges, drumhead surface states~\cite{ChanCa3P2,chiu2018quantized}, and quasitopological electromagnetic responses~\cite{sekinetopocharge,RamamurthytoporesNLSM}. 
The low-energy excitations around the nodal ring of these semimetals are described by one-parameter 
families of (2+1)-dimensional quantum field theories with parity anomalies~\cite{ruitopological}. 
That is, the electromagnetic responses of these nodal rings are given by Chern-Simons actions, which break parity symmetry. 
These Chern-Simons terms lead to transverse Hall effects, where electrons
from opposite sides of the nodal ring flow to opposite surfaces, when an electric field is applied~\cite{ruitopological}. 
Unfortunately, due to time-reversal symmetry, the total current generated by the Chern-Simons action vanishes, once the
sum over all momenta is taken. Therefore, the electric-field induced Hall currents can only be measured
by special devices, that filter electrons based on their momenta~\cite{ruitopological}. 

In this letter, we propose to use pseudo electric fields, induced by lattice vibrations, to probe the parity anomaly of NLSMs. 
As opposed to external electric fields, pseudo electric fields are axial, as they couple with opposite sign to electrons 
with opposite momenta. This permits to directly probe the parity anomaly of NLSMs, via the response
to axial electric fields.  We derive a low-energy description of NLSMs in the presence of strain,
and show that periodic lattice deformations generate a topological piezoelectric effect (TPEE), which originates
from the parity anomaly. This piezoelectric effect manifests itself by dissipationless Hall currents
that are transverse to the deformation field. We show
that the TPEE can be interpreted as a polarization current and that it has a semi-quantized transport coefficient, given by the length of the nodal ring. Furthermore, we discuss experimental considerations for the observation of the TPEE in the 
NLSM materials CaAgP and Ca$_3$P$_2$.

\section{Model}
First, we introduce a lattice model for a NLSM with a single nodal ring, and discuss its topological properties. 
We consider the following tight-binding Hamiltonian on the cubic lattice
\begin{eqnarray}
    {H}({\bm p})&=&t\left[2+\cos p_0a-\cos p_xa-\cos p_ya-\cos p_z a \right]\tau_z \nonumber\\
    &+&v\sin p_z a \, \tau_y +\Delta\tau_x  ,
    \label{eq:lattceham}
\end{eqnarray}
where $\tau_i\;(i=1,2,3)$   are Pauli matrices acting in orbital space.
For simplicity, we assume $t,\;v,\;p_0>0$. 
To discuss the parity anomaly and the electric polarization, we have introduced a small parity-breaking term $\Delta\tau_x$. 
In the absence of  $\Delta\tau_x$, the lattice Hamiltonian is parity-time ($\mathcal{PT}$) symmetric 
 with the $\mathcal{PT}$ operator $\mathcal{PT}=\tau_z K$. 
This tight-binding Hamiltonian describes the low-energy dispersion near the Fermi level of CaAgP and Ca$_3$P$_2$~\cite{yamakageCaAgAs,ChanCa3P2}. 
The symmetry-breaking term $\Delta \tau_x$ can be induced by applying uniaxial pressure, or an electric field~\cite{ruitopological}. 

In the absence of $\Delta \tau_x$, Hamiltonian~\eqref{eq:lattceham} exhibits a nodal ring within the $p_z=0$ plane, centered around $\Gamma$.
This nodal ring is topologically protected by the following $\mathbb{Z}_2$ invariant $\nu[S^1]$~\cite{Zhaosymmetry},
\begin{eqnarray}
    \nu[S^1]=\frac{1}{\pi}\sum_{\alpha\in {\rm occ.\;states}}\oint_{S^1} d{\bm p}  \cdot \mathcal{{\bm A}}^{\alpha,\alpha}({\bm p})\;\;{\rm mod}\;2,
    \label{eq:z2topo}
\end{eqnarray}
where the integration path is along the closed loop $S^1$. Here, $\mathcal{A}^{\alpha,\beta}_i({\bm p}) = i\langle u_\alpha({\bm p})|\partial_{p_i}|u_\beta({\bm p}) \rangle$ and $|u_\alpha ({\bm p}) \rangle$ are the Berry connection and the Bloch eigenstates of Eq.~\eqref{eq:lattceham}, respectively. 
$\mathcal{PT}$ symmetry restricts Eq.~\eqref{eq:z2topo} to the values $\nu[S^1]=0,\;1$~\cite{XiaoBerry}. 
When the loop $S^1$ encircles the nodal ring, we obtain $\nu[S^1]=1$, otherwise $\nu[S^1]=0$.
 
The topological protection of the nodal ring is linked to a bulk electric polarization. 
To see this, let us 
decompose the three-dimensional Hamiltonian~\eqref{eq:lattceham} into one-dimensional subsystems
parametrized by the inplane momenta ${\bm p}_\perp=(p_x, p_y)$.
The electric polarization of each subsystem is given by the Zak's phase~\cite{King-Smithpolarization},
\begin{eqnarray}
    P_z({\bm p}_\perp)&=&\sum_{\alpha\in {\rm occ.\;states}}  \int^{\pi}_{-\pi}\frac{dp_z}{2\pi}\mathcal{A}_z^{\alpha,\alpha}({\bm p})=0,\;\frac{1}{2} ,
    \label{eq:z2polarization}
\end{eqnarray}
and the total polarization is the summation of these phases over the inplane momenta
\begin{eqnarray}
    P_z&=&\int \frac{d{\bm p}_\perp}{(2\pi)^2} P_z({\bm p}_\perp) .
    \label{eq:polarization_w_PTsym}
\end{eqnarray}
From Eqs.~\eqref{eq:z2topo}and~\eqref{eq:z2polarization}, we find that the Zak's phase is $\tfrac{1}{2}$ for a region of inplane momenta ${\bm p}_\perp$ 
that is bounded by the nodal ring. By the bulk-boundary correspondence~\cite{chiu2018quantized}, this leads to midgap surface states at the (001) face of the NLSM, whose 
fillings determine the surface charge.
Due to $\mathcal{PT}$ symmetry, the surface states at the top and bottom (001) faces are degenerate,
thus the electric polarization is determined only up to a multiple of the elementary charge.
To unambiguously determine the bulk polarization, it is necessary to include an infinitesimal $\mathcal{PT}$ symmetry breaking term $\Delta\tau_x$, which opens a bulk gap and removes the degeneracy of the midgap surface states. 
With the inclusion of $\Delta\tau_x$, we find that the bulk polarization is semi-quantized and given by~\cite{RamamurthytoporesNLSM},
\begin{eqnarray}
    &&P_z=\frac{S}{8\pi^2} \mathop{\mathrm{sgn}} (\Delta),
    \label{eq:polarization_wo_PTsym}
\end{eqnarray}
where $S$ is the area encircled by the nodal ring projected onto the surface Brillouin Zone.

\section{Parity anomaly and Chern-Simons action}
\begin{figure}[t]
\includegraphics[width=5.8cm]{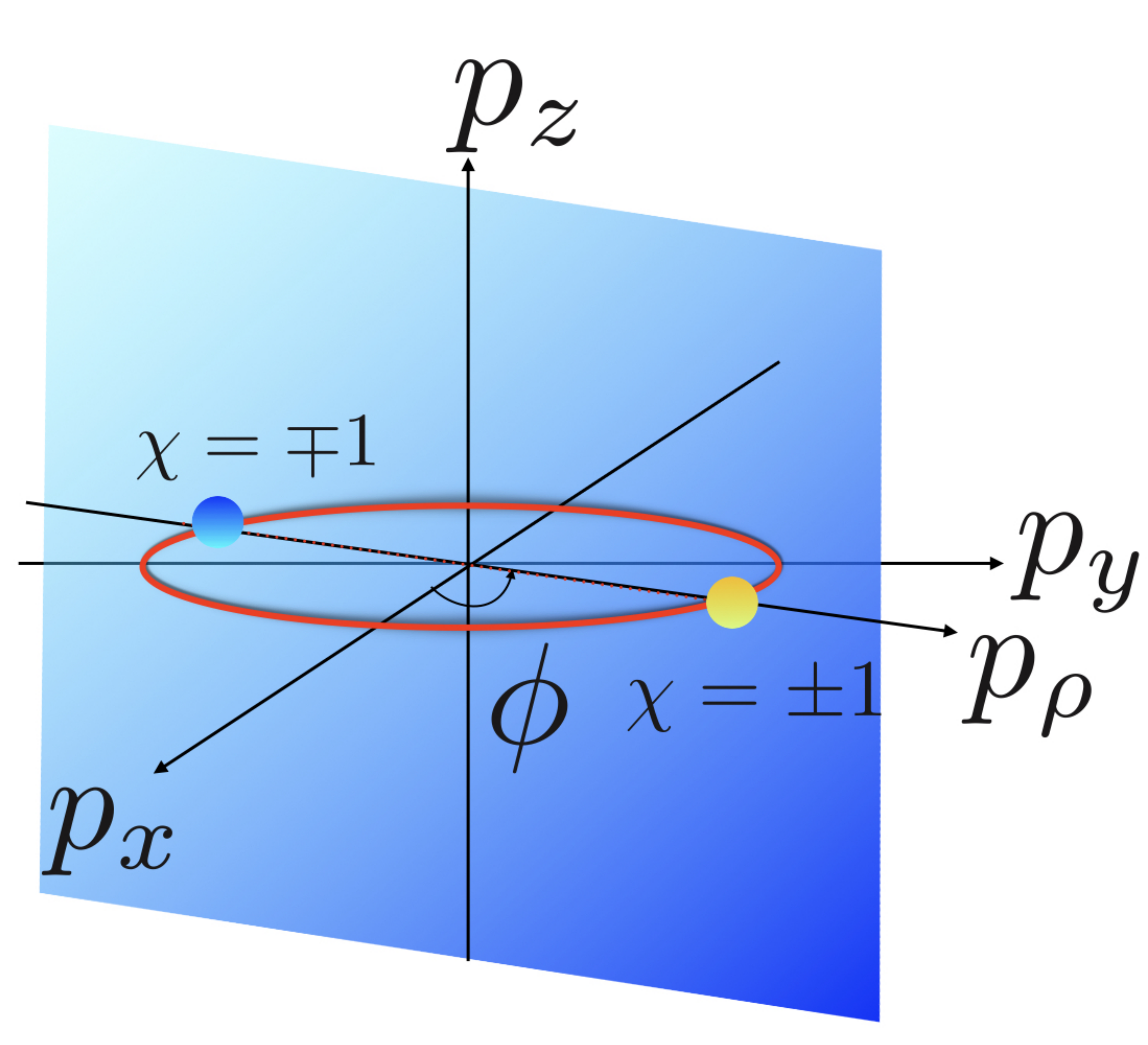}
\caption{
Schematics of the nodal ring (red) within the $p_z=0$ plane. Using the cylindrical coordinates $(p_\rho,\phi,p_z)$, we decompose the NLSM into two-dimensional subsystems (blue) parametrized by $\phi$. Each subsystem contains
two Dirac points with opposite sign of Berry curvature $\chi$. 
}
\label{fig1}
\end{figure}
Next, we use a family of (2+1)-dimensional quantum field theories to
derive the topological responses due to external and pseudo electromagnetic fields.
For small ${\bm p}$, Eq.~\eqref{eq:lattceham} reduces to the 
low-energy continuum Hamiltonian
\begin{eqnarray}
   {H}_{{\rm eff}}({\bm p})&=&\frac{p^2-p_0^2}{2m}\tau_z
    +\lambda p_z \tau_y+\Delta\tau_x,
    \label{eq:effham}
\end{eqnarray}
where $1/(2m) = ta^2/2$ and $\lambda = va$.
Eq.~(\ref{eq:effham}) has rotational symmetry around the $p_z$ axis. 
Thus, we introduce cylindrical coordinates $(p_\rho,\phi,p_z)$ with $p_\rho \in ( -\infty , \infty)$, and  $\phi \in [0,\pi )$, see Fig.~\ref{fig1}. 
With these cylindrical coordinates, we can decompose the three-dimensional system into a family of two-dimensional subsystems  labeled by the azimuth angle $\phi \in [0,\pi)$. 
Each of these subsystems contains two Dirac points that are related by time-reversal symmetry,
and which have opposite sign of Berry curvature $\chi={\rm sign}(\Delta)=\pm1$. 

Each subsystem, labeled by $\phi$, is described by the following (2+1)-dimensional quantum field theory
\begin{eqnarray}
    \label{eq:QFT}
    S^{\phi}&=&\bigoplus_{\chi=\pm1} S^{\phi, \chi}
\\
    S^{\phi,\chi}&=&\int d^2x \, dt \, \overline{\psi}\left[i\chi  \gamma^\mu\left(\partial_\mu+iA_\mu+i\chi A^5_\mu) +\Delta\right)\right]\psi,\nonumber
\end{eqnarray}
where $\psi$ is a two-component Dirac spinor, $\overline{\psi}=\psi^\dag\gamma_0$, $\{ \gamma_\mu,\gamma_\nu \}=\eta_{\mu \nu}$, and $\eta_{\mu \nu}={\rm diag}(1,-1,-1)$. 
The Dirac spinors  interact with the total gauge field $A^\chi_\mu=A_\mu+\chi A^5_\mu$,
which contains both an external gauge field $A_\mu$ and an axial gauge field $A^5_\mu$, respectively.
We note that the axial gauge field couples with opposite sign $\chi$ to the two Dirac points of the subsystem. 
The physical origin of $A^5_\mu$ due to lattice strain will be discussed later. 
Upon regularization~\cite{Redlichparityanomaly}, we obtain from Eq.~\eqref{eq:QFT} the parity breaking Chern-Simons term
\begin{eqnarray}
    S^{\phi,\chi}_{\rm CS}=\frac{\chi}{4\pi}\int d^2x \,dt \,  \epsilon^{\mu \nu \lambda}  A^\chi_\mu \partial_\nu A^\chi_\lambda ,
    \label{eq:CS action}
\end{eqnarray}
which is a manifestation of the parity anomaly.
Varying the Chern-Simons action with respect to $A^\chi_\mu$, gives the
anomalous transverse current
\begin{eqnarray}
    j_\mu^{\phi, \chi}=-\frac{\delta S^{\phi,\chi}_{\rm CS}}{ \delta A_\mu}=\frac{\chi}{4\pi} \epsilon^{\mu \nu \lambda}(\partial_\nu A_\lambda+\chi \partial_\nu A^5_\lambda),
    \label{eq:anomaly}
\end{eqnarray}
for a single Dirac point with chirality $\chi$ in subsystem~$\phi$.
We observe that transverse currents induced by external electromagnetic fields cancel out, since contributions
from opposite sides of the nodal ring have opposite sign $\chi = \pm 1$.
Currents induced by axial gauge fields, however, do not cancel, since they have the same sign everywhere along the nodal ring.
This remarkable feature originates from the axial nature of the strain-induced gauge field $A^5_\mu$,
which couples oppositely to Dirac fermions  with opposite momenta. 

{\section{Strain-induced axial gauge field}
\begin{figure*}
\includegraphics[width=19cm]{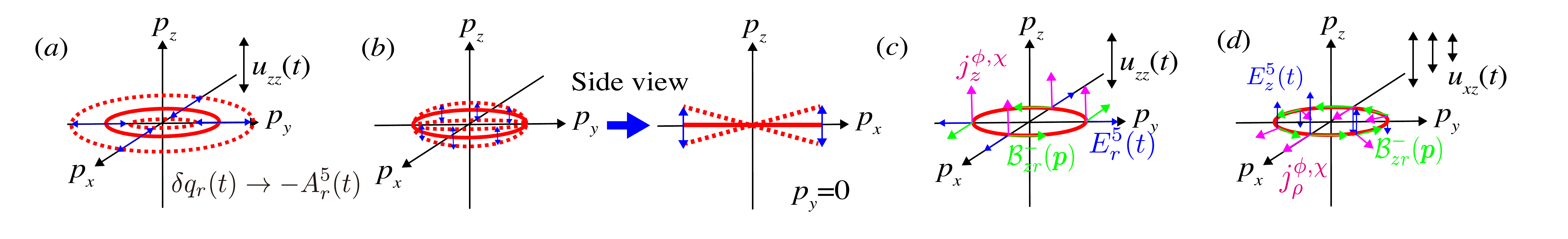}
\caption{(a),(b)  The axial gauge potential $A_r^5(t)$ changes the size of the nodal ring, while $A_z^5(t)$ tilts it out of plane,
as indicated by the dashed lines.
(c),(d)
The strain induced currents ${\bm j}^{\phi, \chi}$ (pink) are perpendicular to both the Berry curvature $\mathcal{B}_{zr}^-({\bm p})$ (green) and the axial electric field ${\bm E}^5$ (blue).
}
\label{fig3}
\end{figure*}
We  now  discuss the physical origin of the axial gauge field. 
The basic idea is to
incorporate lattice strain into the tight-binding model~\eqref{eq:lattceham}, which acts on the low-energy excitations
as effective gauge fields. 
Strain shifts the lattice sites ${\bm R}$ by the displacement vector ${\bm u}({\bm R})$, as ${\bm R}+{\bm u}({\bm R})$,
thereby modifying and introducing new overlaps between atomic orbitals.
In our tight binding model this changes the hopping parameters as~\cite{Cortijostrain,Shapourianstrain}
\begin{subequations} \label{modified_hoppings}
\begin{eqnarray}
\label{eq:hoppingintegralx}
    t({\bm a}_x)\tau_z&\simeq& t(1-u_{xx})\tau_z+ivu_{xz}\tau_y, \\
\label{eq:hoppingintegraly}
    t({\bm a}_y)\tau_z&\simeq& t(1-u_{yy})\tau_z+ivu_{yz}\tau_y, \\
\label{eq:hoppingintegralz}
    t({\bm a}_z)\tau_z&\simeq& t(1-u_{zz})\tau_z, \\
 \label{eq:hoppingintegralzhyb}
    iv ({\bm a}_z)\tau_y&\simeq&  iv(1-u_{zz})\tau_y+t\sum_{i\neq z}u_{zi}\tau_z,
\end{eqnarray}
\end{subequations}
where $ t({\bm a}_\mu) (\mu=x,y,z)$ represents the hopping amplitudes
along the bond direction ${\bm a}_\mu$, and 
$u_{\mu \nu} = (\partial_\mu u_\nu({\bm R})+\partial_\nu u_\mu({\bm R}))/2$ is the symmetrized strain tensor. 
The first terms in Eqs.~\eqref{modified_hoppings} describe changes in the hopping amplitudes between two like orbitals, when the bond lengths are modified by strain.
The second terms originate from new hopping processes between different orbitals, which are symmetry forbidden in the unstrained lattice. 
In the following, we focus on the gauge fields induced by the $u_{z \nu}$ components of the strain tensor, as these are the ones that probe the parity anomaly.
The other components of $u_{\mu \nu}$ only renormalize the Fermi velocity, which is not important for our purpose. 
Using Eq.~\eqref{modified_hoppings} we find that this lattice strain generates additional terms in the tight-binding Hamiltonian~\eqref{eq:lattceham},
$H({\bm p})\to H({\bm p})+\delta H({\bm p})$, which are of the form~\footnote{Here, we have also neglected the term $v u_{zz} \sin k_z \tau_y$, as it
only renormalizes the hopping amplitude $v$.}
\begin{eqnarray}
\delta H({\bm p}) &\simeq& -tu_{zz}\cos p_za \, \tau_{z}
\\
&& \qquad + v(u_{xz}\sin p_xa+u_{yz}\sin p_ya) \tau_y .
\nonumber
    \label{eq:strain}
\end{eqnarray}
These modifications change the low-energy Hamiltonian~\eqref{eq:effham} to 
\begin{subequations}     \label{eq:effham w/ strain}
\begin{eqnarray}
    &&H_{\rm eff}({\bm p})+ \delta H_{\rm eff}({\bm p})   \\
    && \qquad \simeq v_F\left( q_r- A^5_r \right)\tau_z+\lambda (p_z-A^5_z( \varphi )) \tau_y+\Delta\tau_x ,
    \nonumber
\end{eqnarray}
with the pseudo gauge potentials 
\begin{eqnarray} \label{def_axial_gauge_potentials}
 A^5_r &=& u_{zz}/(p_0a^2), 
 \\
 A^5_z( \varphi ) &=& \sum_i f_i( \varphi )A^{i,5}_z\simeq -p_0\left(  u_{xz} \cos \varphi  + u_{yz} \sin \varphi  \right)   ,
\nonumber
\end{eqnarray}
\end{subequations}
along the $r$- and $z$-directions, respectively, where 
$f_x(\varphi )=p_0\cos \varphi$, $f_y( \varphi )=p_0\sin \varphi $, and   $A_z^{i,5}=-u_{iz}\;(i=x,y)$.  
Here, we have introduced the Fermi velocity $v_F = p_0/m$, 
the radial momentum $q_r = p_r-p_0$, and the cylindrical coordinates $(p_r,p_z,\varphi)$ with $p_r\in [0,\infty),\;p_z\in (-\infty,\infty),\; \varphi\in [0,2\pi)$.

We conclude that in NLSMs with a nodal ring in the $p_z =0$ plane, the strain field components $u_{z \nu}$ act on the low-energy
excitations like effective gauge potentials. 
Interestingly, these effective gauge potentials are axial, since they couple with opposite sign to the excitations at opposite sides of the nodal ring, i.e., at $(p_0,0,\varphi)$
and $(p_0,0,\varphi+\pi)$.
From Eq.~\eqref{eq:effham w/ strain}, we see that $A^5_r$ concentrically shrinks or expands the nodal ring, while $A^{i,5}_z$ tilts the nodal ring out of the $p_z=0$ plane, see Figs.~\ref{fig3}(a) and~\ref{fig3}(b), respectively.

If we consider time-dependent lattice strain, i.e., lattice vibrations, we can also generate axial electric fields. 
That is, the time dependence of the strain tensor $u_{\mu z}(t)$ produces axial electric fields via
\begin{eqnarray}
E_r^5 
 &=&
-\frac{\partial A_r^5 }{\partial t},
\qquad
E_z^{i,5} 
=
-\frac{\partial A_z^{i,5} }{\partial t},
\end{eqnarray}
where the axial electric fields $E_z^{i,5}$ are defined by the angnular independent parts of the axial vector potentials, and $f_i(\varphi)$
are absorbed into the axial charge coupling constants.

\section{Topological piezoelectric effect}
Next, we demonstrate that axial electric fields in NLSMs induce
net topologic currents that flow in the direction perpendicular to the axial fields.
For that purpose we use linear response theory to compute the
axial conductivity tensors  
 $\tilde{\sigma}_{\mu r}(\omega)$ and $\tilde{\sigma}^x_{\mu r}(\omega)$, which are defined as
\begin{eqnarray}
    \langle \hat{j}_\mu \rangle(\omega)=\tilde{\sigma}_{\mu r}(\omega)E^5_r(\omega)+\tilde{\sigma}^x_{\mu z}(\omega)E^{x,5}_z(\omega) ,
\end{eqnarray}
with the current density operator $\hat{{\bm j}}$.
By use of Kubo's formula we compute the axial Hall conductivities $\tilde{\sigma}_{z r}(T,\omega)$ and $\tilde{\sigma}_{x z}(T,\omega)$.
In the DC limit $\omega\to 0$, they are given by
\begin{subequations} \label{eq_DC_Hall_conduct}
\begin{eqnarray}
    \tilde{\sigma}_{z r}^{\rm DC}(T)&=&-\frac{1}{V}\sum_{\bm p,\alpha}f(\epsilon^\alpha_{\bm p} ) \mathcal{B}^\alpha_{zr}({\bm p}),
    \label{DC axial conductivity zr} \\
    \tilde{\sigma}_{x z}^{x,\rm DC}(T)&=&-\frac{1}{V}\sum_{\bm p,\alpha}f(\epsilon^\alpha_{\bm p} )p_0\cos \varphi \mathcal{B}^\alpha_{xz}({\bm p}),
    \label{DC axial conductivity zr}
\end{eqnarray}
\end{subequations}
where $\epsilon^\alpha_{\bm p}$, $f(\epsilon^\alpha_{\bm p} )$ and $\mathcal{B}^\alpha_{\mu \nu}({\bm p})=-2 {\rm Im} \langle \partial_{p_\mu} u_\alpha({\bm p})|\partial_{p_\nu}u_\alpha({\bm p}) \rangle$ are the energy of the Bloch electrons, the Fermi function, and the Berry curvature, respectively~\footnote{See the Appendix A for details of the derivation of Eq.~(15) and a discussion of the effects of spin-orbit coupling.}.

Thus, it follows that axial electric fields produce transverse Hall currents, whose magnitude is 
determined by the Berry curvature. These Hall currents are perpendicular to both the axial electric field
and the Berry curvature, see Figs.~\ref{fig3}(c) and~\ref{fig3}(d). 
For instance, axial electric fields along the $r$ direction lead to electric currents in the $z$ direction,
since the direction of the Berry curvature is within the $p_z=0$ plane.
Similarly, axial electric fields along the $z$ direction produce currents in the $p_z=0$ plane.
Because the axial electric fields are generated by lattice vibrations, we refer to this type of Hall response as {\it a topological piezoelectric effect}. 

Interestingly, in the low-frequency regime $|\omega/\Delta| \ll 1$, the axial Hall conductivities become semi-quantized, 
i.e., their magnitude depends only on the length of the nodal ring $L=2\pi p_0$. That is, in
the limit $|\omega/\Delta| \ll 1$ we find
\begin{subequations}
\begin{eqnarray}
    \tilde{\sigma}_{z r}^{\rm DC}(T=0)\simeq-\frac{L}{8\pi^2}{\rm sign}(\Delta),\\
    \label{quantizedDC_conductivity_zr}
    \tilde{\sigma}_{x z}^{x,\rm DC}(T=0)\simeq \frac{L}{16\pi^2}{\rm sign}(\Delta),
    \label{quantizedDC axial conductivity}
\end{eqnarray}
\end{subequations}
where $|\Delta / (v_F p_{\rm cut}) |,\; |\Delta/ (\lambda p_{\rm cut}) |\ll 1$ is assumed,
with some cut-off momentum $p_{\rm cut}$.
This is confirmed by numerical evaluations
of $\tilde{\sigma}_{zr}(T=0,\omega)$, see Fig.~\ref{fig4}. 
We observe  in Figs.~\ref{fig4}(a) and~\ref{fig4}(b) 
that the axial Hall conductivity 
 asymptotically approaches its semi-quantized value for $\omega \to 0$,
once $\Delta$ becomes sufficiently small and $p_{\rm cut}$ sufficiently large, respectively. 
As displayed in the inset of Fig.~\ref{fig4}(a) the semi-quantized value of the DC axial Hall conductivity scales linearly with the size of the nodal ring.

\begin{figure}[b]
\includegraphics[width=8.6cm]{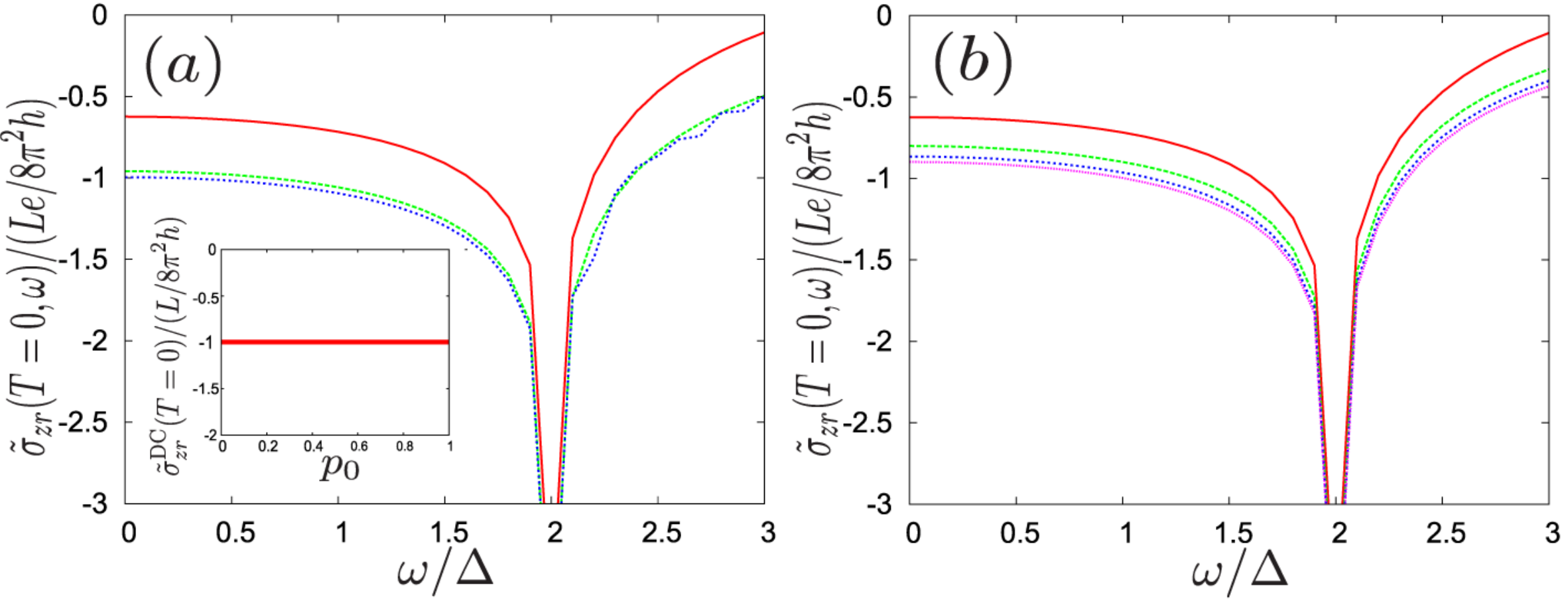}
\caption{
Frequency dependence of the axial Hall conductivity $\tilde{\sigma}_{zr}(T=0,\omega)$
for different values of $\Delta$ and $p_{\rm cut}$.
In (a) the red, green, and blue curves correspond to $\Delta/(v_Fp_0)=0.1$, $0.01$, and $0.001$, respectively,
with $p_{\rm cut}$ fixed at $p_{\rm cut}=0.2 p_0$.
In (b) the red, green, blue, and pink curves correspond to 
$p_{\rm cut} /( p_0 / 10 ) =2$, $4$, $6$, and $8$, respectively, with 
$\Delta $ fixed at $\Delta =0.1 v_Fp_0$.
The inset shows the $p_0$ dependence of the DC axial Hall conductivity $\tilde{\sigma}^{\rm DC}_{zr}(T=0)$
with $p_{\rm cut}=0.2p_0$ and $\Delta=0.0004$.
The other parameters in all panels are: $v_F=2.0$, $p_0=0.2$, and $\lambda=2.5$.  }
\label{fig4}
\end{figure}

Before concluding, we show that the TPEE is related  to the polarization current of NLSMs. 
As discussed above, the axial electric field $E_r^5(t)$ periodically shrinks and expands
the nodal ring. 
This leads to a periodic fluctuation of the bulk electric polarization, which is determined by the size of the nodal ring.
Hence, the axial electric field generates a polarization current, which according to Eq.~\eqref{eq:polarization_wo_PTsym}, takes the form
\begin{eqnarray}
j_{z}^{\rm pol}=\frac{dP_z(t)}{dt}=\frac{1}{8\pi^2}\frac{dS(t)}{dt}\simeq -\frac{L}{8\pi^2}{\rm sign}(\Delta)E_r^5(t),
\label{pol_current}
\end{eqnarray}
where $S(t)=\pi (p_0+A_r^5(t))^2$ is the area of the nodal ring.  
Since Eq.~\eqref{pol_current} coincides with Eq.~(\ref{quantizedDC_conductivity_zr}),
we conclude that the TPEE is linked to the polarization current of NLSMs and that its quantization arises from the semi-quantized electric polarization.

\begin{figure}[t]
\includegraphics[width=8cm]{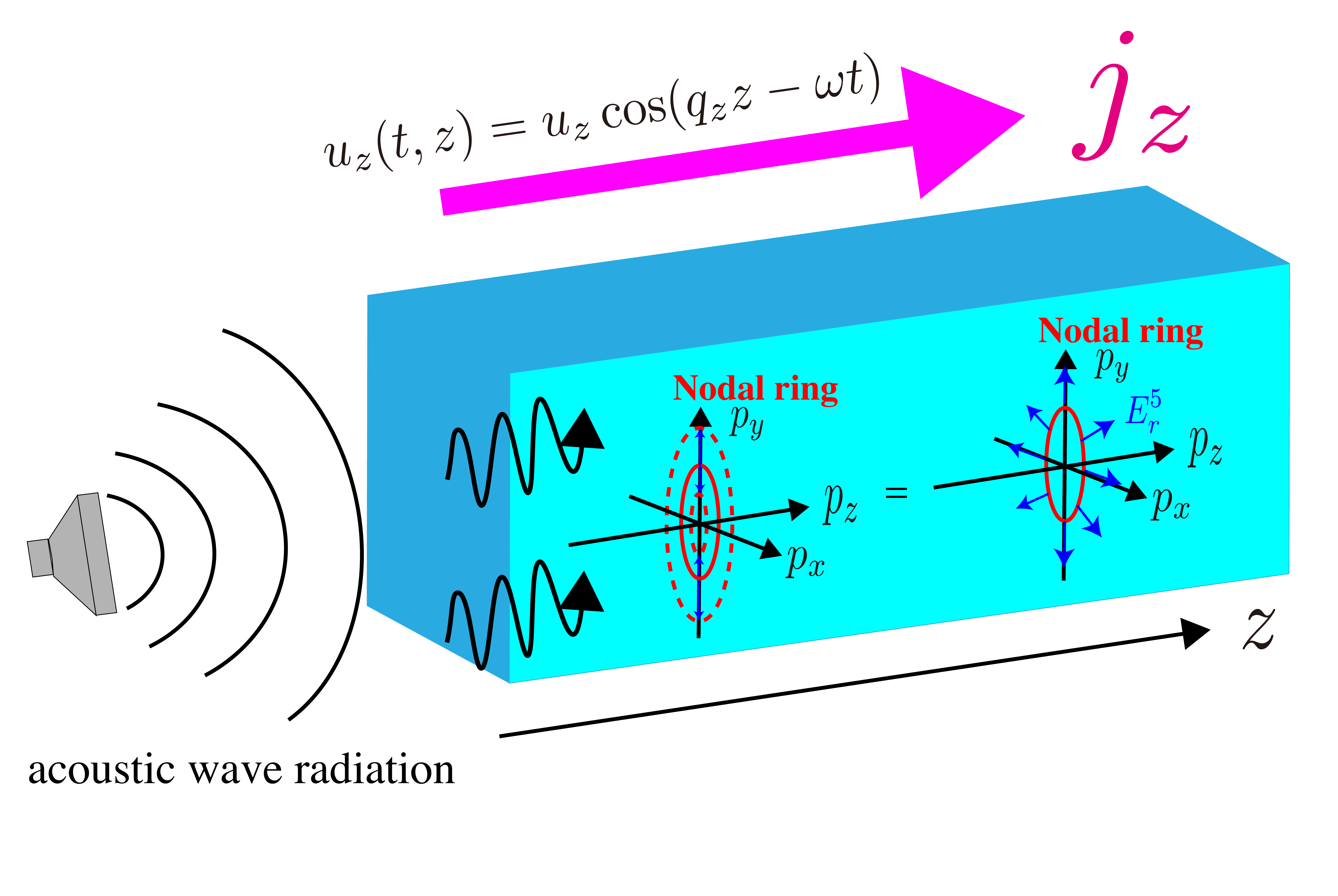}
\caption{
Schematic of the experimental setup for the detection of the TPEE induced by ultrasonic wave radiation.  
}
\label{fig_TPEE}
\end{figure}

The proposed TPEE  is testable in the materials CaAgP or Ca$_3$P$_2$, which possess
a single nodal ring near the Fermi energy.
In these materials lattice vibrations can be generated by sound waves, which leads to an AC current via the TPEE (see Fig.~\ref{fig_TPEE})~\cite{sukhachov2020acoustogalvanic,ferreiros2019mixed}. 
Radiating an ultrasonic wave along the $z$-direction generates a displacement vector of the form
\begin{eqnarray}
u_{z}(t,z)=u_{z} \cos (q_z z-\omega t),
\label{u_z}
\end{eqnarray}
where $q_z$ is the wave vector and $\omega$ is the frequency of the acoustic wave.
From Eq.~(\ref{u_z}) the amplitude of the strain tensor is roughly given by $\sim u_{z}q_z/(2\pi)$.
To estimate the typical magnitude of the TPEE in the case of Ca$_3$P$_2$, we consider
an ultrasonic wave with the frequency $\omega=100\;{\rm MHz}$,   
$q_z=2.0\times 10^4\;{\rm m}^{-1}$, and sound velocity $\sim$~5000$~{\rm m/sec}$.
Further, we assume that $u_{z}=5.0\;{\rm \AA}$ and the parity breaking term is about $|\Delta/(v_Fp_0)|=1.0 \times 10^{-3}$.
The relevant material parameters for Ca$_3$P$_2$ are $v_F=2.72\times 10^5\; {\rm m/s},\; \lambda
=3.80\times 10^5\; {\rm m/s},\; p_0=0.206\; {\rm \AA}^{-1}$, and $a=8.26\;{\rm \AA}$~\cite{ChanCa3P2}. 
From this we estimate the TPEE current to be of about $j_z\simeq 1.9\; {\rm mA}/{\rm cm}^2$, which is experimentally detectable~\footnote{As verified in Ref.~\cite{yamakageCaAgAs} and \cite{ChanCa3P2}, the spin-orbit couplings in CaAgP and Ca$_3$P$_2$ are negligible. Also see Supplemental Materials on effects of the spin-orbit couplings on the TPEE.}.

\section{Conclusion} 
We have shown that dynamical strain induces a TPEE in NLSMs,
which manifests itself by dissipationless Hall currents, originating from the parity anomaly.
We have focused on NLSMs with negligible spin-orbit coupling (SOC);
the case of strong SOC is briefly discussed in~\cite{com0}. 
While it would be of fundamental interest to observe the parity anomaly in NLSMs using sound waves, the 
TPEE could also be useful for future piezoelectric devices, e.g., sound wave or vibration detectors.
Since NLSMs can be easily switched between a semimetallic regime with high mobility and 
a piezoelectric regime using, e.g., strain, they could also be used for controllable multi-functional devices.
Finally, we note that the concept of the TPEE can be extended straightforwardly  to 2D Dirac semimetals.

\section*{Acknowlegement} 
T.M.\ thanks K. Nomura, A. Yamakage, and Y. Ominato for invaluable discussion. T.M.\ was supported by a JSPS Fellowship for Young Scientists. S.F.\ was supported by the Grant-in-Aids for Scientific
Research from JSPS of Japan (Grants No. 17K05517), and KAKENHI on Innovative Areas ``Topological Materials Science'' (No.~JP15H05852) and "J-Physics" (No.~JP18H04318), and JST CREST Grant Number JPMJCR19T5, Japan.
This research was initiated at the KITP UC Santa Barbara and supported in part by the National Science Foundation under Grant No.~NSF PHY-1748958.


\begin{appendix}

\section{Derivation of Eq.~(15)}


To derive the axial Hall conductivities $\tilde{\sigma}_{z r}$ and $\tilde{\sigma}^x_{x z}$ we
use the Bloch eigenbasis  $\{ | u_{\pm} ( {\bm p} ) \rangle \}$ with
\begin{equation}
\begin{aligned}
    &H_{\rm eff}({\bm p})|u_\pm({\bm p}) \rangle=\epsilon^\pm_{\bm p} |u_\pm({\bm p}) \rangle,\\
    &\epsilon^\pm_{\bm p} = \pm \sqrt{v_F^2q_r^2+\lambda^2p_z^2+\Delta^2},
\end{aligned}
\end{equation} 
and express the current density operator as
\begin{equation}
    \hat{{\bm j}} = \sum_{{\bm p},\alpha=\pm}\frac{\partial \epsilon^\alpha_{\bm p}}{\partial {\bm p}}\hat{c}_{{\bm p} \alpha}^\dag \hat{c}_{{\bm p} \alpha}
     + i\sum_{\alpha \neq \beta}\left(\epsilon^\alpha_{\bm p}-\epsilon^\beta_{\bm p} \right)
    \mathcal{{\bm A}}^{\alpha,\beta}({\bm p})\hat{c}_{{\bm p} \alpha}^\dag \hat{c}_{{\bm p} \beta},
\end{equation}
where  $\hat{c}_{{\bm p} \alpha}^\dag$ ($\hat{c}_{{\bm p} \alpha}$) are creation (annihilation) operators.

From this expression and with the use of the Kubo's formula we obtain  the axial Hall conductivities,
\begin{subequations} \label{eq_full_Hall_conduct}
\begin{eqnarray}
    \label{sig_zr}
 &&\tilde{\sigma}_{z r}(T,\omega)
    =
    -\lim_{\delta \to+0}\frac{i}{V}
    \sum_{\bm p,\alpha \neq \beta} f(\epsilon^\alpha_{\bm p})(\epsilon^\alpha_{\bm p}-\epsilon^\beta_{\bm p}) \nonumber\\
 && \qquad \times 
    \left[ 
    \frac{\mathcal{A}_z^{\alpha ,\beta}\mathcal{A}_r^{\beta ,\alpha}}{\omega+\epsilon^\alpha_{\bm p}-\epsilon^\beta_{\bm p} +i\delta}
    +\frac{\mathcal{A}_r^{\alpha ,\beta}\mathcal{A}_z^{\beta , \alpha}}{\omega-\epsilon^\alpha_{\bm p}+\epsilon^\beta_{\bm p}+i\delta}  
\right] \\
    \label{sig_xz}
&&\tilde{\sigma}^x_{x z}(T,\omega)
    =
    -\lim_{\delta \to+0}\frac{i}{V}
    \sum_{\bm p,\alpha \neq \beta} f(\epsilon^\alpha_{\bm p}) p_0 \cos \varphi (\epsilon^\alpha_{\bm p} - \epsilon^\beta_{\bm p})  \nonumber\\
&& \qquad \times  
   \left[ 
        \frac{\mathcal{A}_x^{\alpha ,\beta}\mathcal{A}_z^{\beta ,\alpha}}{\omega+\epsilon^\alpha_{\bm p}-\epsilon^\beta_{\bm p}+i\delta}
        +\frac{\mathcal{A}_z^{\alpha ,\beta}\mathcal{A}_x^{\beta ,\alpha}}{\omega-\epsilon^\alpha_{\bm p}+\epsilon^\beta_{\bm p}+i\delta} 
\right] ,
\end{eqnarray}
\end{subequations}
We note that the factor $p_0 \cos \varphi$ in Eq.~(\ref{sig_xz}) originates from the momentum dependence of the axial electric field.
In the DC limit $\omega\to 0$, we obtain Eq.~(15) in the main text,
\begin{subequations} \label{eq_DC_Hall_conduct}
\begin{eqnarray}
    \tilde{\sigma}_{z r}^{\rm DC}(T)&=&-\frac{1}{V}\sum_{\bm p,\alpha}f(\epsilon^\alpha_{\bm p} ) \mathcal{B}^\alpha_{zr}({\bm p}),
    \label{DC axial conductivity zr} \\
    \tilde{\sigma}_{x z}^{x,\rm DC}(T)&=&-\frac{1}{V}\sum_{\bm p,\alpha}f(\epsilon^\alpha_{\bm p} )p_0\cos \varphi \mathcal{B}^\alpha_{xz}({\bm p}),
    \label{DC axial conductivity zr}
\end{eqnarray}
\end{subequations}
where $\mathcal{B}^\alpha_{\mu \nu}({\bm p})=-2 {\rm Im} \langle \partial_{p_\mu} u_\alpha({\bm p})|\partial_{p_\nu}u_\alpha({\bm p}) \rangle$ is the Berry curvature, which is given by, 
\begin{subequations}
\begin{eqnarray}
    \mathcal{B}^\pm_{zr}({\bm p})&=&\mp  \frac{\Delta \lambda v_F}{2(v_F^2q_r^2+\lambda^2p_z^2+\Delta^2)^{3/2}},\\
    \mathcal{B}^\pm_{xz}({\bm p})&=&\pm \frac{\Delta \lambda v_F\cos \varphi}{2\left(v_F^2q_r^2+\lambda^2p_z^2+\Delta^2\right)^{3/2}}.
    \label{Berry curvature}
\end{eqnarray}
\end{subequations}

\section{Effects of spin orbit coupling}

In the main text we have considered materials with very weak SOC, such as Ca$_3$P$_2$ and CaAgP, where
the topological properties are protected by SU(2) spin-rotation symmetry in combination with reflection symmetry and/or $\mathcal{PT}$ symmetry.
In these materials SOC is at least three orders of magnitude smaller than the band width. Such small SOC slightly breaks the
SU(2) spin-rotation symmetry and possibly also the reflection symmetry, which opens a very small gap at the nodal line.
Nevertheless, such small SOC does not substantially alter the Berry curvature carried by the bands. As a result the transverse topological currents and the TPEE
remain largely unchanged, only their magnitude is slightly reduced and the semi quantization slightly broken~\cite{ChanCa3P2}. 
In Ca$_3$P$_2$ and CaAgP SOC is so small that these effects can be neglected. 
Hence, these materials are good platforms for the experimental verification of the TPEE.

However, there are also nodal-line materials with strong SOC, which fully breaks the SU(2) spin-rotation symmetry. This either splits the Dirac nodal-line into
two Weyl nodal lines or leads to a full gap in the spectrum. The former occurs in PbTaSe$_2$ and TlTaSe$_2$, which exhibit Weyl nodal lines that are protected
by reflection symmetry and time-reversal symmetry~\cite{ChanCa3P2,bianARPES,bian_TlTaSe2_PRB}. 
The topology of these two materials is characterized by a reflection invariant that guarantees the stability of the
nodal line. Moreover, these two materials exhibit a reflection anomaly, that leads to a similar topological
response and TPEE as in CaAgAs and Ca$_3$P$_2$. 

\begin{figure}[t!]
\includegraphics[width=9cm]{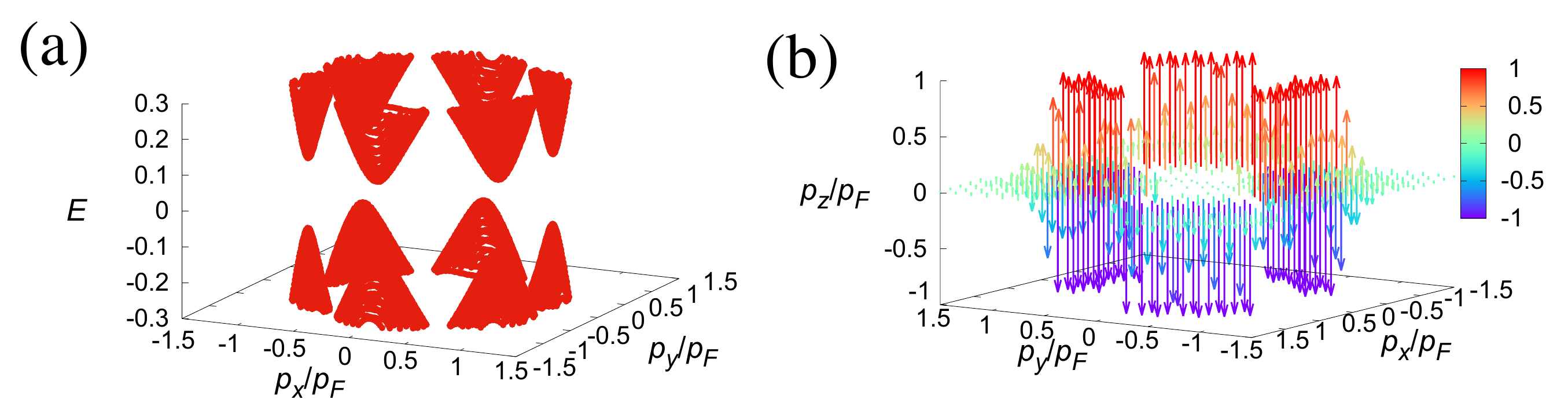}
\caption{(a) Energy dispersion near the nodal line of CaAgAs. SOC opens a gap of about 20~meV.
(b) The Berry curvature of the occupied states of CaAgAs in the presence of SOC.}
\label{figsup}
\end{figure}

An example of a nodal-line material where SOC generates a large gap  is CaAgAs. 
In the absence of SOC, this material has a nodal ring within the $p_z=0$ plane, which is protected by  mirror reflection symmetry.
The SOC in CaAgAs breaks not only SU(2) spin-rotation symmetry, but also the mirror reflection symmetry, which opens up an energy gap of about 20~meV. 
This transforms the nodal-line semimetal into a  topological insulator with a $\mathbb{Z}_2$ invariant~\cite{yamakageCaAgAs}.
We will now study whether the presence of an SOC-induced gap of 20~meV removes the TPEE  in CaAgAs or not. 
For this purpose we consider the following effective continuum Hamiltonian~\cite{yamakageCaAgAs}
\begin{eqnarray}
    H_{\rm CaAgAs}(\bm p)=
    \begin{pmatrix}
    h(\bm p)&&\Lambda(\bm p)\\
    \Lambda^{\dag}(\bm p)&&h^*(-\bm p)
    \end{pmatrix},
    \label{Hamiltonian_CaAgAs}
\end{eqnarray}
where
\begin{eqnarray}
    h(\bm p)&=&\frac{\bm p^2 -p_F^2}{2m}\sigma_z+Ap_z(p_x\sigma_x-p_y\sigma_y),\\
 \label{SOC_term_CaAgAs}   
    \Lambda(\bm p)&=&-iB_\perp p_z(1-\sigma_z)-iB_\parallel (p_x+ip_y)\sigma_x\nonumber\\
    &-&iD\left(p_x-ip_y \right)^2\sigma_y.
    \label{Hamiltonian_CaAgAs}
\end{eqnarray}
Eq.~\eqref{SOC_term_CaAgAs} represent an asymmetric SOC which breaks with its strong momentum dependence the mirror reflection symmetry.
As a consequence, a strongly momentum dependent gap is opened, see Fig.~\ref{figsup}(a). 
The strongly momentum dependent SOC leads to a Berry curvature profile, which is qualitatively different from that
of Ca$_3$P$_2$ (compare Fig.~\ref{figsup}(b) with~Fig.~2 in the main text).
If an axial electric field is applied along the $r$-direction, the topological Hall current in CaAgAs   therefore cancels out. 
Hence, the TPEE in CaAgAs is destroyed by the asymmetric SOC.
This is in agreement with the fact that CaAgAs does not exhibit a semi-quantized bulk electric polarization.
The key for the realization of a TPEE is a semi-quantized bulk electric polarization.
If a large enough SOC breaks it, then the TPEE is not realized.

\end{appendix}

\bibliography{nodalline_ref.bib}
\end{document}